\documentclass[amsmath,amssymb,showkeys,superscriptaddress,aps,prd,10pt,twocolumn]{revtex4-2}
\usepackage{graphicx}
\usepackage[utf8]{inputenc}
\usepackage[caption=false]{subfig}
\usepackage{multirow}
\usepackage{appendix}
\usepackage{graphicx,epstopdf}

\def\xslash{x\!\!\!\slash }

\def\vel{\left|}
\def\ver{\right|}
\usepackage{colortbl}
\usepackage{xcolor}
\usepackage[colorlinks=true, urlcolor=blue, linkcolor=blue, citecolor=blue]{hyperref}

\begin{document}

\title{Magnetic moments of the doubly charged axial-vector \texorpdfstring{$T_{cc}^{++}$}{}  states}
\author{Ula\c{s}~\"{O}zdem}%
\email[]{ulasozdem@aydin.edu.tr}
\affiliation{ Health Services Vocational School of Higher Education, Istanbul Aydin University, Sefakoy-Kucukcekmece, 34295 Istanbul, Turkey}

 
\begin{abstract}
Motivated by the discovery of the doubly-charmed state $T^+_{cc}$ and with the help of light-cone sum rules, the magnetic moments of possible $T_{cc}^{++}$ states are calculated.
While calculating the magnetic moments of these states, these particles are considered in the molecular picture and they have $J^P = 1^+$ quantum numbers. The magnetic moment results obtained for the $T_{cc}^{++}$ states are large due to the double electric charge. The results obtained in this study can be checked using other theoretical models. The magnetic moments of the hadrons reveal valuable knowledge about the size and the shape of the hadrons. Measurement of the magnetic moment of the $T_{cc}^{++}$ states in future experimental collaborations can be very useful to understanding substructure and identification the quantum numbers of these states.
\end{abstract}

\maketitle

\section{Introduction}\label{introduction}

Apart from the traditional hadron state, mesons and baryons, it is theoretically possible in states that contain more quarks. These particles, which cannot fit into the conventional hadron category, are called exotic states. In the last few decades, numerous exotic particles have been observed by different experimental collaborations, and the number of new states reported continues to increase. These newly discovered states not only create excitement in the scientific community, but also bring new questions such as the internal structure of these states and quantum numbers. Many different models have been suggested to elucidate and unravel the nature of these states, and many studies are being conducted on them~\cite{Chen:2016qju,Ali:2017jda,Olsen:2017bmm,Lebed:2016hpi,Guo:2017jvc,Nielsen:2009uh,Brambilla:2019esw,Liu:2019zoy, Agaev:2020zad, Dong:2021juy}.
It should be noted here that most of the exotic states reported in the experimental facilities so far have hidden-charm or hidden-bottom quark structure, i.e., they include $ c \bar c $ or  $ b \bar b $ in their substructures.

Recently, the LHCb Collaboration reported the first observation of doubly-charmed tetraquark state $ T_{cc}^{+}$ in the $D^0 D^0 \pi^+$ invariant mass spectrum~\cite{LHCb:2021vvq,LHCb:2021auc}. 
 The Breit-Wigner parameters of this doubly charmed tetraquark state are $\delta m=-273 \pm 61 \pm 5_{-14}^{+11}~\mathrm{KeV}$ and $\Gamma =410 \pm 165 \pm 43_{-38}^{+18}~\mathrm{KeV}$, where the mass difference is with respect to the $D^0  D^{\ast +}$ threshold. 
The fact that it has an electrical charge and contains two charm quarks makes it a good candidate for the exotic state with a quark content of $c c\bar u \bar d$.
Also, the spin-parity quantum number of this state was estimated by the experiment as $J^P = 1^+$. The importance of the reported $T^+_{cc}$ is identical to that of the X(3872) therefore the  newly discovered state offers a new and important platform for both experimental and theoretical hadron physics.
Before and after the discovery of the $T^+_{cc}$ state, many theoretical studies have been performed to understand spectroscopic properties,  the decay modes/production mechanism and magnetic dipole moments within different configurations~\cite{Qin:2020zlg,Feijoo:2021ppq,Dong:2021bvy,Yan:2021wdl,Wang:2021yld,Huang:2021urd,Agaev:2021vur, Azizi:2021aib,Chen:2021tnn,Jin:2021cxj,Ling:2021bir,Hu:2021gdg,Chen:2021cfl,Albaladejo:2021vln,Abreu:2021jwm,Du:2021zzh,Dai:2021vgf,Wang:2021ajy,Meng:2021jnw,Fleming:2021wmk,Xin:2021wcr,Ren:2021dsi,Albuquerque:2022weq}. In addition to the $T_{cc}$ states the properties of doubly-bottom tetraquark states have also been investigated in different theoretical models~\cite{Ren:2021dsi,Albuquerque:2022weq,Dai:2021wxi,Karliner:2017qjm,Eichten:2017ffp,Cheng:2020wxa,Braaten:2020nwp, Meng:2020knc,Dias:2011mi,Navarra:2007yw,Gao:2020ogo,Agaev:2020mqq,Agaev:2020zag,Agaev:2020dba,Agaev:2019lwh,Agaev:2018khe,Aliev:2021dgx}.
Most of the theoretical studies have obtained the quantum numbers of the observed $T^+_{cc}$ state as $I(J^P) = 0(1^+)$. In the case the $T^+_{cc}$ has $I = 1$, in Ref.~\cite {Albaladejo:2021vln} they predict the location of the other two members ($T^{++}_{cc}$ and $T^{0}_{cc}$) of the triplet. Besides, using Heavy-Quark Spin Symmetry, they predict the location of possible heavier $D^*D^*$ (I = 0 or $I = 1$) partners. 
In Ref.~\cite{Xin:2021wcr}, the isospin  $I=0,1/2$ and $1$ of the $T_{cc}$ states have been investigated by means of the QCD sum rule. In this study, strong and electromagnetic decay channels are also proposed for tetraquarks with these different isospin states.
In Ref.~\cite{Ren:2021dsi}, the masses of the doubly-heavy tetraquark states without strangeness and; with single and double strangeness in the framework of the one-meson exchange potential model. In this study, they predicted that $T_{bb}$ states are more stable than $T_{cc}$ states.  
In Ref.~\cite{Cheng:2020wxa}, the masses of the $QQ\bar q\bar q$ tetraquark states with the help of heavy diquark-antiquark symmetry and the
chromomagnetic interaction model. They predicted that only $bb \bar q \bar q$ and $bb \bar q \bar s$ states are stable  with respect to the strong decays. They also discussed the constraints on the masses of the these tetraquark states.
In Ref.~\cite{Braaten:2020nwp}, they have predicted masses of the doubly-heavy tetraquark states in the heavy quark limit. They found that only doubly-bottom tetraquarks with the $\bar u \bar d$, $\bar s \bar u$ and $\bar s \bar d$ are stable with the strong decays.
In Ref.~\cite{Agaev:2018vag}, the mass and residues of the possible $ T_{cc}^{++}$ tetraquark states were investigated with quantum numbers $J^P =0^-$ within the QCD sum rules method.

 The electromagnetic form factors and multi-pole moments of hadrons, in addition to their spectroscopic properties, may give a hint us identify their exact character, inner structure, and quantum numbers. 
 We know that the electromagnetic multi-pole moments particularly magnetic moment of the hadrons, which contain knowledge about the spatial distributions of charge and magnetization inside the hadrons, are directly associated to the spatial distributions of quarks and gluons inside hadrons. 
 Therefore it is interesting to investigate the magnetic and higher  multi-pole moments of hadrons.
Motivated by the reasons given above, in this study, we evaluate the magnetic moments of possible $ T_{cc}^{++}$  states  in the molecular configuration employing the light-cone sum rule method \cite{Chernyak:1990ag, Braun:1988qv, Balitsky:1989ry}. 
Particles carrying a double-electric charge constitute another very interesting class of exotic states, since the doubly-charged states cannot be described as standard mesons. The doubly-charged states may exist as doubly-charmed tetraquarks  consisted of  the heavy diquark $cc$ and light antidiquarks $\bar s \bar s$ or $\bar d \bar s$. To put it another way, they can include two or three quark flavors. The diquark $bb$ and antidiquark $\bar u \bar u$ can also bind to form the doubly-charged state $ T_{bb}^{--}$ containing only two quark spices. In this study, only the magnetic moments of $ T_{cc}^{++}$ states will be examined.

 This manuscript is organized as follows. After the introduction in Sec.~\ref{introduction},  in Sec.~\ref{formalism} we present the light-cone sum rule formalism and and in this formalism we determine the tools necessary to calculate the magnetic moments of the $T_{cc}^{++}$ states. In the Sec.~\ref{numerical}, we make numerical calculations of magnetic moments using the analytical expressions obtained in the previous section and we discuss these results.
 
 \begin{widetext}

 \section{Formalism} \label{formalism}
 
 In the light-cone sum rule approach, we evaluate a correlation function, as building block of the method,  once  with respect to the hadronic quantities such as coupling constants, form factors, and second, with respect to the QCD parameters and distribution amplitudes of the photon, which are available with regard to different twists. Then desirable physical quantity obtained by equating the coefficients of appropriate Lorentz structures from both representations of the correlation function and employing the assumption of quark-hadron ansatz. 
 
 %
In order to calculate magnetic moment in light cone sum rules, we need a correlation function in an external electromagnetic field and its written as follows 

\begin{equation}
 \label{edmn01}
\Pi _{\mu \nu }(p,q)=i\int d^{4}xe^{ip\cdot x}\langle 0|\mathcal{T}\{J_{\mu}^{i}(x)
J_{\nu }^{i \dagger }(0)\}|0\rangle_{\gamma}, 
\end{equation}%
where  the $\gamma$ denotes the external electromagnetic field and $J_{\mu}^i(x)$ is the interpolating current of the $T_{cc}^{++}$ states with the spin-parity quantum numbers $ J^{P} = 1^{+}$.  In this study, since $T_{cc}^{++}$ states are examined in molecular configuration, molecular forms of interpolation currents are needed and those currents are written as follows~\cite{Xin:2021wcr}

\begin{eqnarray}\label{curr}
   J_{\mu}^1(x)&= |DD^*\rangle_{I=1} &= \frac{1}{\sqrt{2}}\Big\{\big[\bar d_a(x) i\gamma_5 c_a(x)][\bar d_b(x) \gamma_\mu c_b(x)]+[\bar d_a(x) \gamma_\mu c_a(x)\big]\big[\bar d_b(x)  i\gamma_5 c_b(x)\big]\Big\},\nonumber\\
     J_{\mu}^2(x)&= |DD_s^*\rangle_{I=1/2} &=  \frac{1}{\sqrt{2}}\Big\{\big[\bar d_a(x) i\gamma_5 c_a(x)\big]\big[\bar s_b(x) \gamma_\mu c_b(x)\big]+\big[\bar d_a(x) \gamma_\mu c_a(x)][\bar s_b(x)  i\gamma_5 c_b(x)\big]\Big\},
    \nonumber \\
     J_{\mu}^3(x)&= |D_sD_s^*\rangle_{I=0} &= \Big\{\big[\bar s_a(x) i\gamma_5 c_a(x)\big]\big[\bar s_b(x) \gamma_\mu c_b(x)\big]\Big\},\nonumber\\
     \nonumber\\
      J_{\mu}^4(x)&= |D_1D_0^*\rangle_{I=1} &= \frac{1}{\sqrt{2}}\Big\{\big[\bar d_a(x)  c_a(x)\big]\big[\bar d_b(x)  \gamma_\mu \gamma_5 c_b(x)\big]+\big[\bar d_a(x) \gamma_\mu c_a(x)\big]\big[\bar d_b(x)  i\gamma_5 c_b(x)\big]\Big\},\nonumber\\
     J_{\mu}^5(x)&= |D_{s1}D_0^*\rangle_{I=1/2} &= \frac{1}{\sqrt{2}}\Big\{\big[\bar d_a(x)  c_a(x)\big]\big[\bar s_b(x)  \gamma_\mu \gamma_5 c_b(x)\big]+\big[\bar d_a(x) \gamma_\mu c_a(x)\big]\big[\bar s_b(x)  i\gamma_5 c_b(x)\big]\Big\},
    \nonumber \\
    J_{\mu}^6(x)&= |D^*_{s1}D_{s0}^*\rangle_{I=0} &= \Big\{\big[\bar s_a(x)  c_a(x)\big]\big[\bar s_b(x)  \gamma_\mu \gamma_5 c_b(x)\big]\Big\}.
     \end{eqnarray}


 In the hadronic representation, the correlation function is obtained by inserting a full set of hadronic state and integrating the four-dimensional x. After these processes, the correlation function is written as
 
\begin{align}
\label{edmn04}
\Pi_{\mu\nu}^{Had} (p,q) = {\frac{\langle 0 \mid J_\mu (x) \mid
T_{cc}^{++}(p, \varepsilon^\theta) \rangle}{p^2 - m_{T_{cc}^{++}}^2}} \langle T_{cc}^{++}(p, \varepsilon^\theta) \mid T_{cc}^{++}(p+q, \varepsilon^\delta) \rangle_\gamma
\frac{\langle T_{cc}^{++}(p+q,\varepsilon^\delta) \mid {J^\dagger}_\nu (0) \mid 0 \rangle}{(p+q)^2 - m_{T_{cc}^{++}}^2} + \cdots,
\end{align}
where  q is momentum of the  photon and dots represent the effects of the higher states and continuum. The matrix elements in Eq. (\ref{edmn04}) 
are written as
\begin{align}
\label{edmn05}
\langle 0 \mid J_\mu^{T_{cc}^{++}}(x) \mid T_{cc}^{++}(p,\varepsilon^\theta) \rangle &= \lambda_{T_{cc}^{++}} \varepsilon_\mu^\theta\,,
\\
\nonumber\\
\langle T_{cc}^{++}(p+q,\varepsilon^\delta) \mid {J^\dagger}_\nu (0) \mid 0 \rangle&=\lambda_{T_{cc}^{++}} \varepsilon_\nu^\delta\,,
\\
\nonumber\\
\langle T_{cc}^{++}(p,\varepsilon^\theta) \mid  T_{cc}^{++} (p+q,\varepsilon^{\delta})\rangle_\gamma &= - \varepsilon^\tau (\varepsilon^{\theta})^\alpha (\varepsilon^{\delta})^\beta \Big\{ G_1(Q^2)~ (2p+q)_\tau ~g_{\alpha\beta}  + G_2(Q^2)~ ( g_{\tau\beta}~ q_\alpha -  g_{\tau\alpha}~ q_\beta) \nonumber\\ &- \frac{1}{2 m_{T_{cc}^{++}}^2} G_3(Q^2)~ (2p+q)_\tau ~q_\alpha q_\beta  \Big\},\label{edmn06}
\end{align}
where $\lambda_{T_{cc}^{++}}$, $ \varepsilon_\mu^\theta\ $ and $ \varepsilon_\nu^\delta\ $  are represent the residue, initial and final  polarization vectors of the $T_{cc}^{++}$  states, respectively. Here,  $G_1(Q^2)$, $G_2(Q^2)$ and $G_3(Q^2)$ are Lorentz invariant form factors,  with  $Q^2=-q^2$.

The final form of the correlation function in the hadronic representation is obtained using Eqs. (\ref{edmn04})-(\ref{edmn06}) as follows:
\begin{align}
\label{edmn09}
 \Pi_{\mu\nu}^{Had}(p,q) &=  \frac{\varepsilon_\rho \, \lambda_{T_{cc}^{++}}^2}{ [m_{T_{cc}^{++}}^2 - (p+q)^2][m_{T_{cc}^{++}}^2 - p^2]}
 \Big\{ G_2 (Q^2) \Big(q_\mu g_{\rho\nu} - q_\nu g_{\rho\mu} -
\frac{p_\nu}{m_{T_{cc}^{++}}^2}  \big(q_\mu p_\rho - \frac{1}{2}
Q^2 g_{\mu\rho}\big) 
 + \nonumber\\
 &  +
\frac{(p+q)_\mu}{m_{T_{cc}^{++}}^2}  \big(q_\nu (p+q)_\rho+ \frac{1}{2}
Q^2 g_{\nu\rho}\big)
-  
\frac{(p+q)_\mu p_\nu p_\rho}{m_{T_{cc}^{++}}^4} \, Q^2
\Big)
\nonumber\\
&
+\mbox{other independent structures}\Big\}\,+\cdots.
\end{align}

Only the value of the $G_2(Q^2)$ form factor at $Q^2 = 0$ is required to define the magnetic moment. 
The magnetic form factor $F_M(Q^2)$ is defined as 
\begin{align}
\label{edmn07}
&F_M(Q^2) = G_2(Q^2)\,,
\end{align}
 the magnetic form factor $F_M(Q^2=0)$, is proportional to the
 magnetic moment $\mu_{T_{cc}^{++}}$ :
\begin{align}
\label{edmn08}
&\mu_{T_{cc}^{++}} = \frac{ e}{2\, m_{T_{cc}^{++}}} \,F_M(0).
\end{align}

In the second step of magnetic moment calculation of $T_{cc}^{++}$ states, correlation function is calculated with respect to the QCD's degrees of freedom and photon distribution amplitudes.
For this purpose, in QCD representation, we contract the related quark fields by means of Wick's theorem after substituting the explicit expressions of the interpolating currents in the correlation function. As an example, we give the result of $J_\mu^3$ current as follows

\begin{eqnarray}
\Pi _{\mu \nu }^{\mathrm{QCD}}(p,q)&=&i\int d^{4}xe^{ip\cdot x} \langle 0 \mid \Bigg\{  \mathrm{%
Tr}\Big[ \gamma _{5} S_{c}^{aa^{\prime }}(x)\gamma
_{5}S_{s}^{a^{\prime }a}(-x)\Big]    
\mathrm{Tr}\Big[ \gamma _{\mu }S_{c}^{bb^{\prime
}}(x)\gamma _{\nu }S_{s}^{bb^{\prime }}(-x)\Big]\notag \\
&&+ \mathrm{%
Tr}\Big[ \gamma _{5} S_{c}^{ab^{\prime }}(x)\gamma
_{\nu}S_{s}^{b^{\prime }a}(-x)\Big]    
\mathrm{Tr}\Big[ \gamma _{\mu }S_{c}^{ba^{\prime
}}(x)\gamma _{5 }S_{s}^{a^{\prime }b}(-x)\Big]\notag \\
&& -\mathrm{Tr}\Big[ \gamma
_{5}S_{c}^{ab^{\prime }}(x) \gamma _{\nu}S_{s}^{b^{\prime }b}(-x)  \gamma _{\mu }S_{c}^{ba^{\prime }}(x)\gamma _{5}S_{s}^{a^{\prime }a}(-x)\Big]\notag\\
&& -\mathrm{Tr}\Big[ \gamma
_{5}S_{c}^{aa^{\prime }}(x) \gamma _{5}S_{s}^{a^{\prime }b}(-x)  \gamma _{\mu }S_{c}^{bb^{\prime }}(x)\gamma _{\nu}S_{s}^{b^{\prime }a}(-x)\Big]\Bigg\} \mid 0 \rangle_{\gamma} ,  \label{eq:QCDSide}
\end{eqnarray}%
 where $S_{q}(x)$ and $S_{c}(x)$ are represent the full light and charm quark propagators.  
Throughout our calculations, we use the x-space expressions for the light and charm quark propagators:

\begin{align}
\label{edmn12}
S_{q}(x)&=i \frac{{\xslash}}{2\pi ^{2}x^{4}} 
- \frac{\langle \bar qq \rangle }{12} \Big(1-i\frac{m_{q} \xslash}{4}   \Big)
- \frac{ \langle \bar qq \rangle }{192}m_0^2 x^2  \Big(1-i\frac{m_{q} \xslash}{6}   \Big)
-\frac {i g_s }{32 \pi^2 x^2} ~G^{\mu \nu} (x) \Big[\rlap/{x} 
\sigma_{\mu \nu} +  \sigma_{\mu \nu} \rlap/{x}
 \Big],
\end{align}
\begin{align}
\label{edmn13}
S_{c}(x)&=\frac{m_{c}^{2}}{4 \pi^{2}} \Bigg[ \frac{K_{1}\Big(m_{c}\sqrt{-x^{2}}\Big) }{\sqrt{-x^{2}}}
+i\frac{{\xslash}~K_{2}\Big( m_{c}\sqrt{-x^{2}}\Big)}
{(\sqrt{-x^{2}})^{2}}\Bigg]
-\frac{g_{s}m_{c}}{16\pi ^{2}} \int_0^1 dv\, G^{\mu \nu }(vx)\Bigg[ \big(\sigma _{\mu \nu }{\xslash}
  +{\xslash}\sigma _{\mu \nu }\big)\nonumber\\
  &\times \frac{K_{1}\Big( m_{c}\sqrt{-x^{2}}\Big) }{\sqrt{-x^{2}}}
+2\sigma_{\mu \nu }K_{0}\Big( m_{c}\sqrt{-x^{2}}\Big)\Bigg],
\end{align}%
where $\langle \bar qq \rangle$ is quark  condensate, $m_0$ is characterized via the quark-gluon mixed condensate  $\langle 0 \mid \bar  q\, g_s\, \sigma_{\alpha\beta}\, G^{\alpha\beta}\, q \mid 0 \rangle = m_0^2 \,\langle \bar qq \rangle $, $G^{\mu\nu}$ is the gluon field strength tensor, and $K_1$, $K_2$ and $K_3$ are modified Bessel functions of the second kind. 
The first term of the light and heavy quark propagators correspond to perturbative or free part and the rest belong to the interacting parts. 
%

The correlation function in Eq. (\ref{eq:QCDSide}) include different contributions: the photon can be emitted both perturbatively or non-perturbatively. In the first situation, the photon interacts with one of the light or heavy quarks, perturbatively. In this case,  
the propagator of the quark interacting with the photon perturbatively is modified via
\begin{align}
\label{free}
S^{free}(x) \rightarrow \int d^4y\, S^{free} (x-y)\,\rlap/{\!A}(y)\, S^{free} (y)\,,
\end{align}
the remaining three propagators in Eq.~(\ref{eq:QCDSide}) are replaced with the full quark propagators involving the perturbative and the non-perturbative contributions. 
In the second situation, one of the light quark propagators in Eq.~(\ref{eq:QCDSide}), defined the photon emission at large distances, is replaced via
\begin{align}
\label{edmn14}
S_{\mu\nu}^{ab}(x) \rightarrow -\frac{1}{4} \big[\bar{q}^a(x) \Gamma_i q^b(x)\big]\big(\Gamma_i\big)_{\mu\nu},
\end{align}
and the other  propagators are replaced with the full quark propagators.
 Here, $\Gamma_i$ are the full set of Dirac matrices. Once 
Eq. (\ref{edmn14}) is plugged into Eq. (\ref{eq:QCDSide}), there appear matrix
elements like $\langle \gamma(q)\vel \bar{q}(x) \Gamma_i q(0) \ver 0\rangle$
and $\langle \gamma(q)\vel \bar{q}(x) \Gamma_i G_{\mu\nu}q(0) \ver 0\rangle$, denoting the non-perturbative contributions. 
These matrix elements are parameterized in connection with distribution amplitudes of the photon, which were determined in Ref. \cite{Ball:2002ps}.
The QCD degrees of freedom representation of the correlation function can be obtained associated with the quark-gluon parameters via  distribution amplitudes of the photon and after performing an integration over x, the expression of the correlation function in the momentum representation can be achieved.

As a final step, the $q_\mu \varepsilon_\nu$ structure is selected from both representations and the coefficients of the structure are matched from both the hadronic and QCD representations. To remove the effects of the continuum and higher states, Borel transformation and continuum subtraction are applied. These steps are routine and quite lengthy for the light-cone sum rules method, so we will not give these technical analyzes here. Interested readers can find technical details on these applications in Ref.~\cite{Azizi:2018duk}. 
As an example, we give the result of $J_\mu^3$ current as follows

\begin{align}
 &\mu_{J_\mu^3}\,\, \lambda_{J_\mu^3}^2  = e^{\frac{m_{J_\mu^3}^2}{M^2}} \,\, \Pi^{QCD}.
\end{align}

The results obtained for the $\Pi^{QCD}$ function are given in Appendix.
Analytical calculations of magnetic moments of $T_{cc}^{++}$ states end here. In the next section, we will perform numerical calculations using these analytical results.

\end{widetext}

\section{Numerical illustrations and conclusions}\label{numerical}

In the present section, we perform a numerical calculations of the light-cone sum rules for the magnetic moments of the $T^{++}_{cc}$ states.
We use $m_u=m_d=0$, $m_s =96^{+8}_{-4}\,\mbox{MeV}$,
$m_c = (1.275\pm 0.025)\,$GeV, 
$\langle \bar ss\rangle $= $0.8 \langle \bar uu\rangle$ with 
$\langle \bar uu\rangle $=$(-0.24\pm0.01)^3\,$GeV$^3$~\cite{Ioffe:2005ym},  
$m_0^{2} = 0.8 \pm 0.1$~GeV$^2$~\cite{Ioffe:2005ym} and $\langle \frac{\alpha_s}{\pi} G^2 \rangle =(0.012\pm0.004)$ $~\mathrm{GeV}^4 $~\cite{Belyaev:1982cd}. 
Numerical values of the mass and residue parameters of these $T_{cc}^{++}$ states are also required in order to perform a numerical analysis of the magnetic moment of these $T_{cc}^{++}$ states. These values were calculated in Ref. \cite{Xin:2021wcr} with the help of mass sum rules and these results are used in the numerical analysis of the text.
The wave functions in the  distribution amplitudes of the photon and all associated expressions are taken from Ref.~\cite{Ball:2002ps}.

\begin{widetext}
 
 \begin{figure}[htp]
\centering
\subfloat[]{\includegraphics[width=0.45\textwidth]{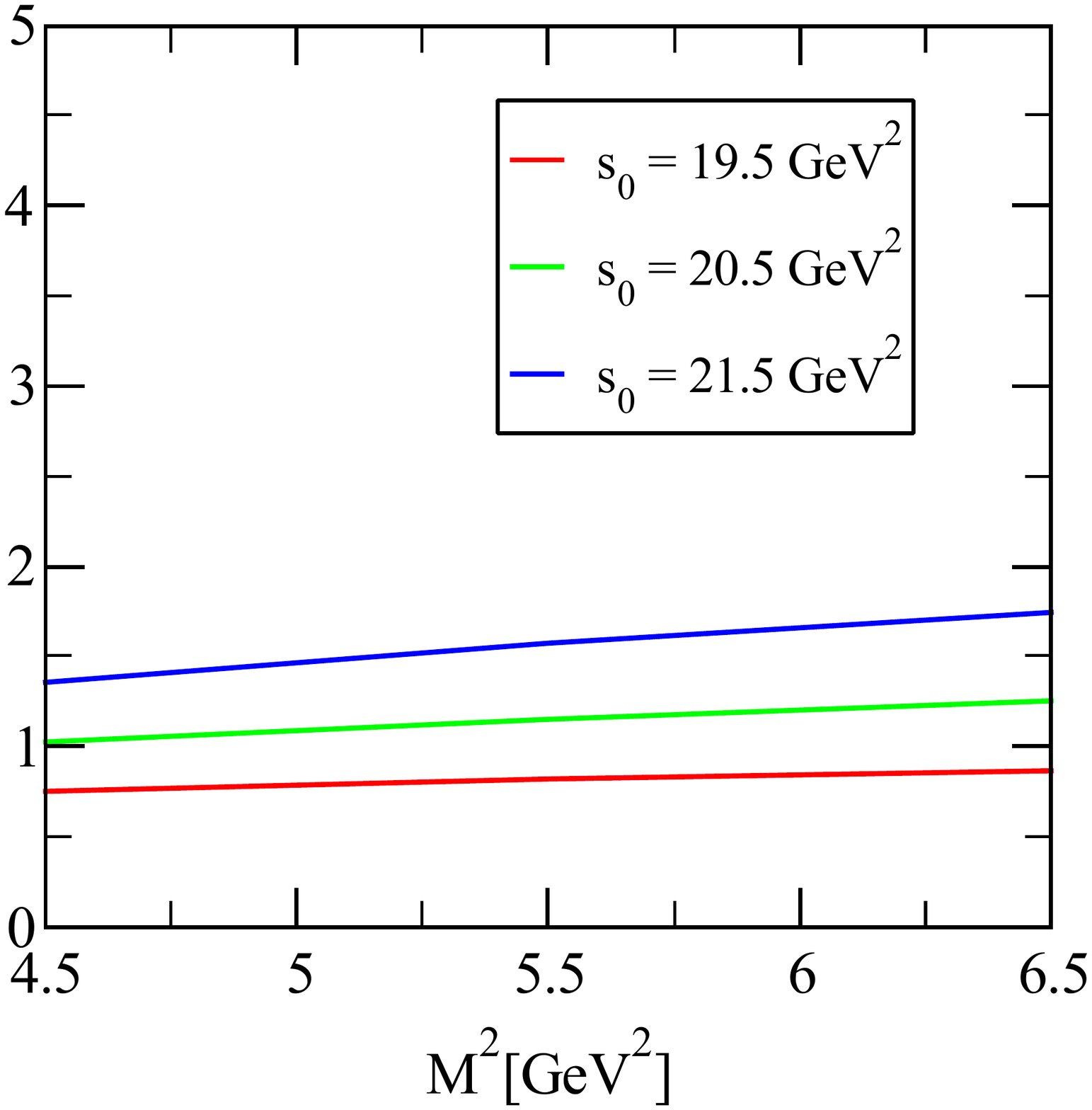}}
\subfloat[]{\includegraphics[width=0.45\textwidth]{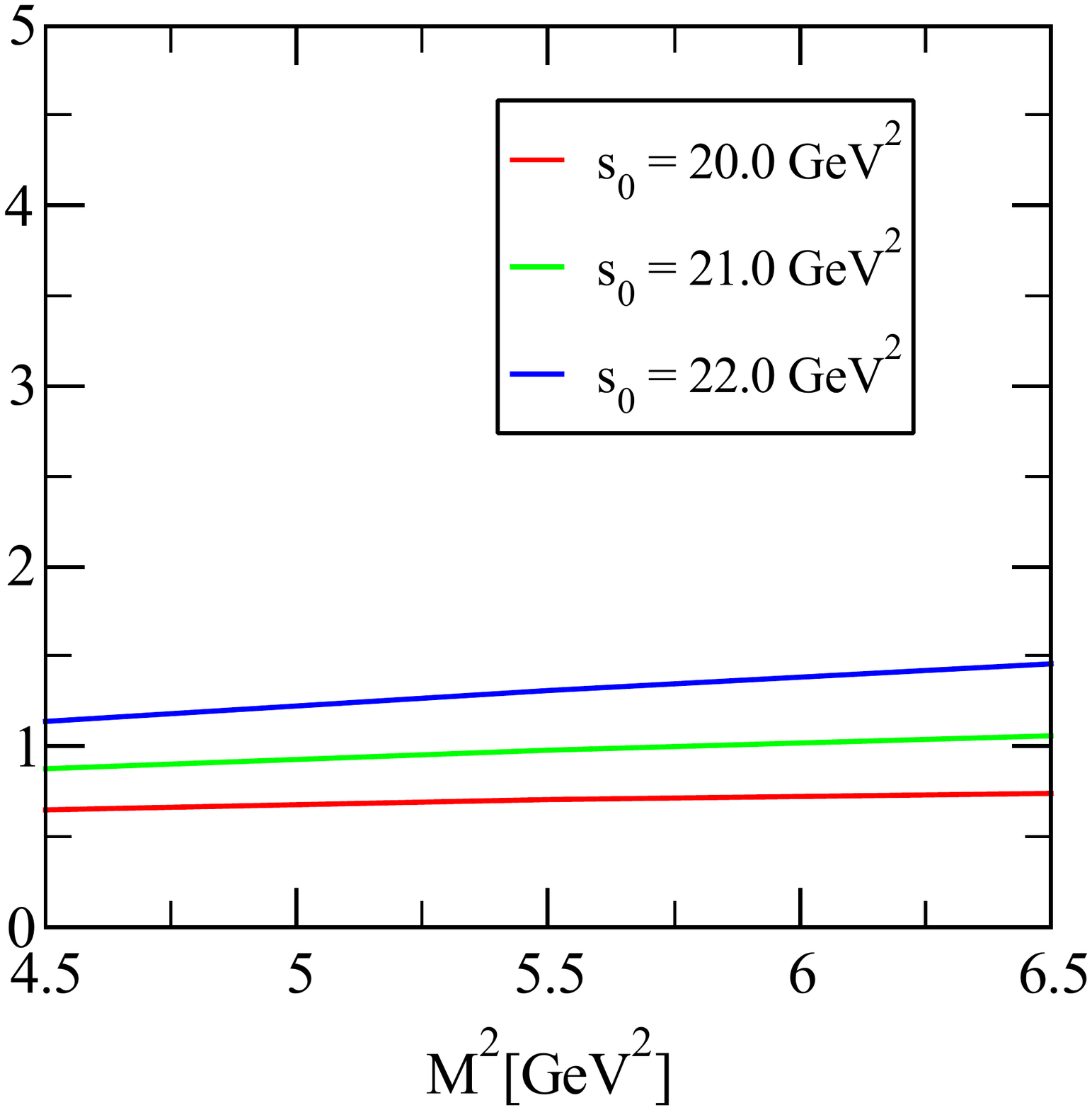}}\\ 
\subfloat[]{\includegraphics[width=0.45\textwidth]{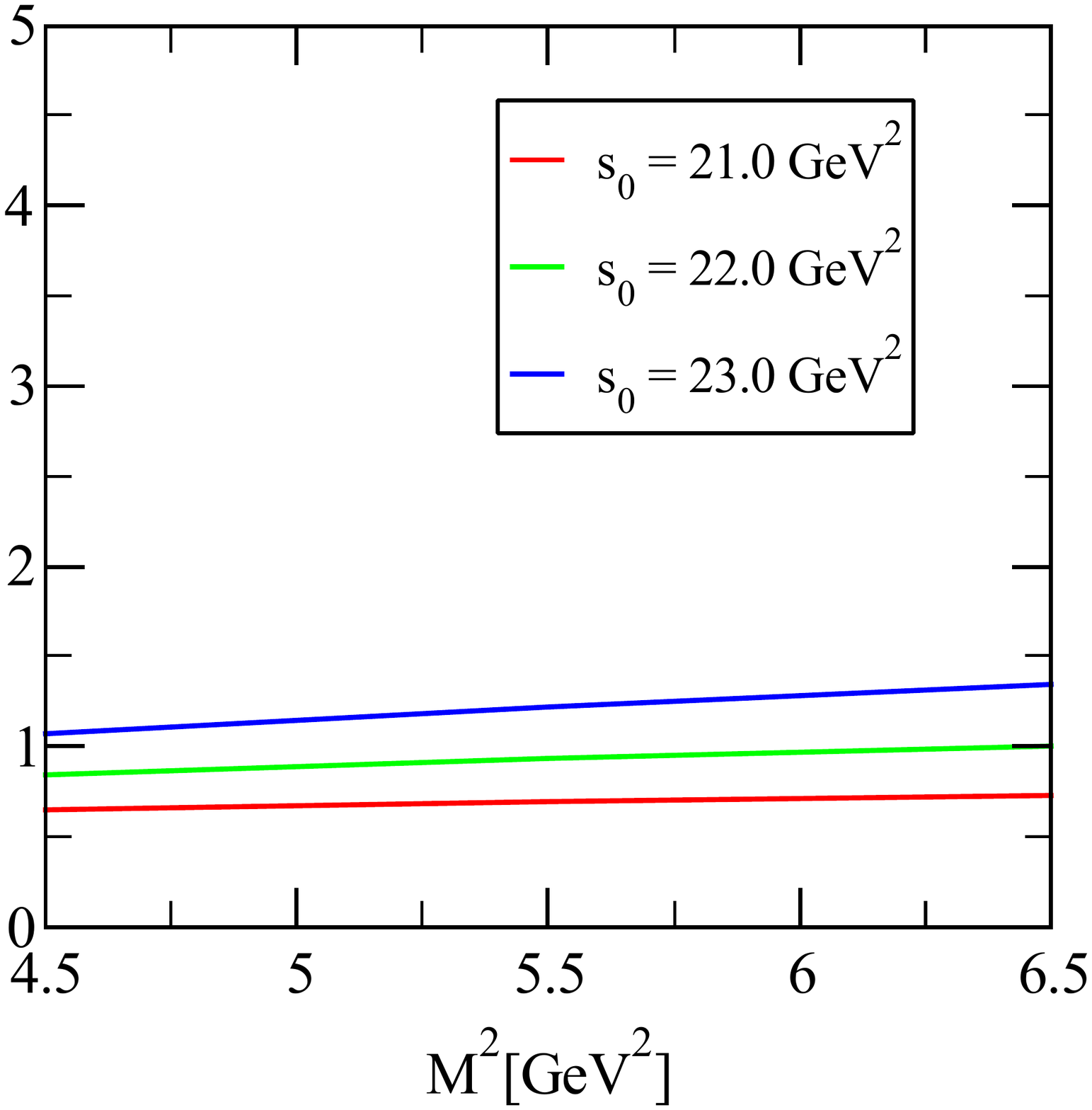}}
\subfloat[]{\includegraphics[width=0.45\textwidth]{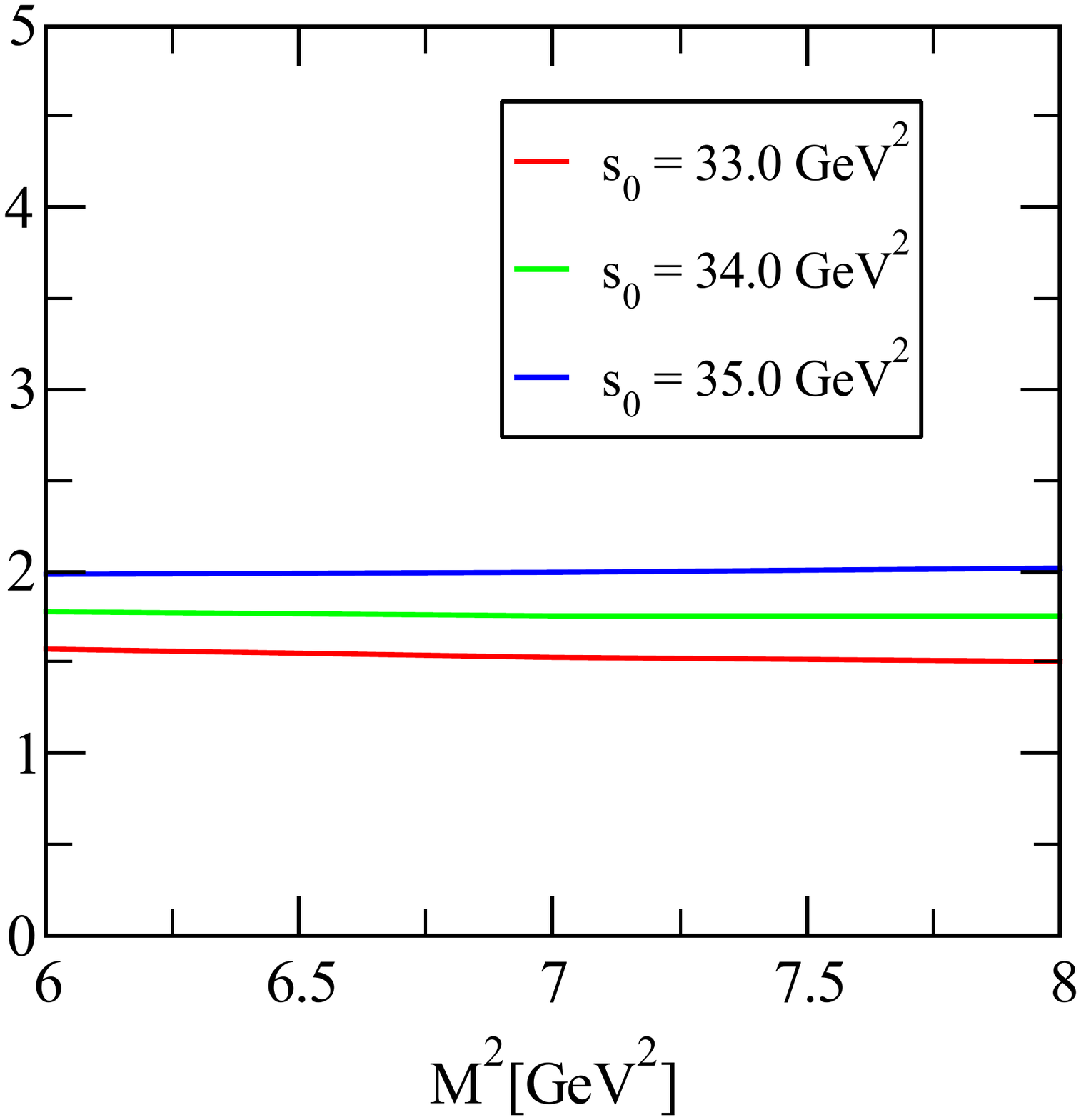}}\\
\subfloat[]{\includegraphics[width=0.45\textwidth]{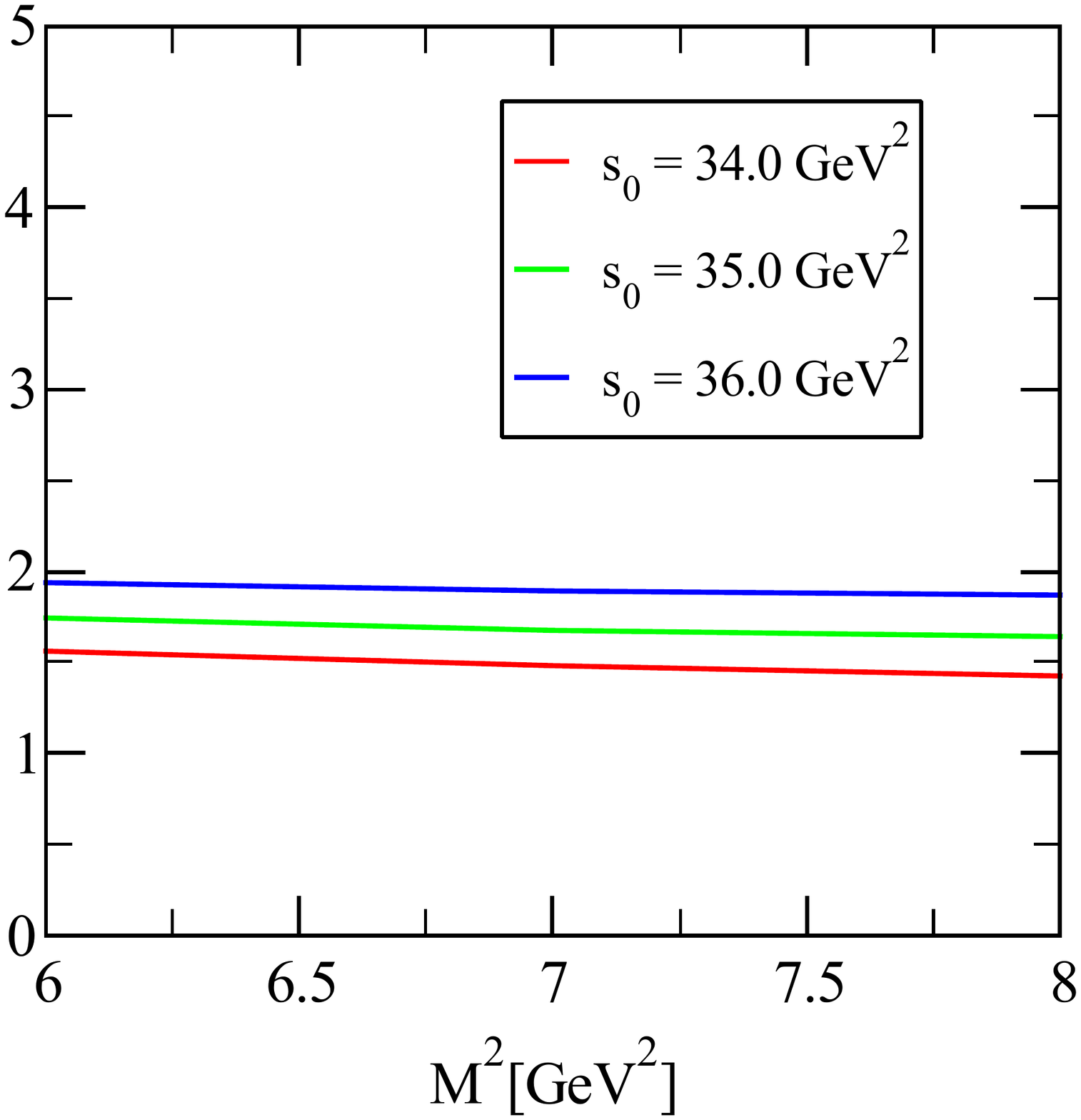}}
\subfloat[]{\includegraphics[width=0.45\textwidth]{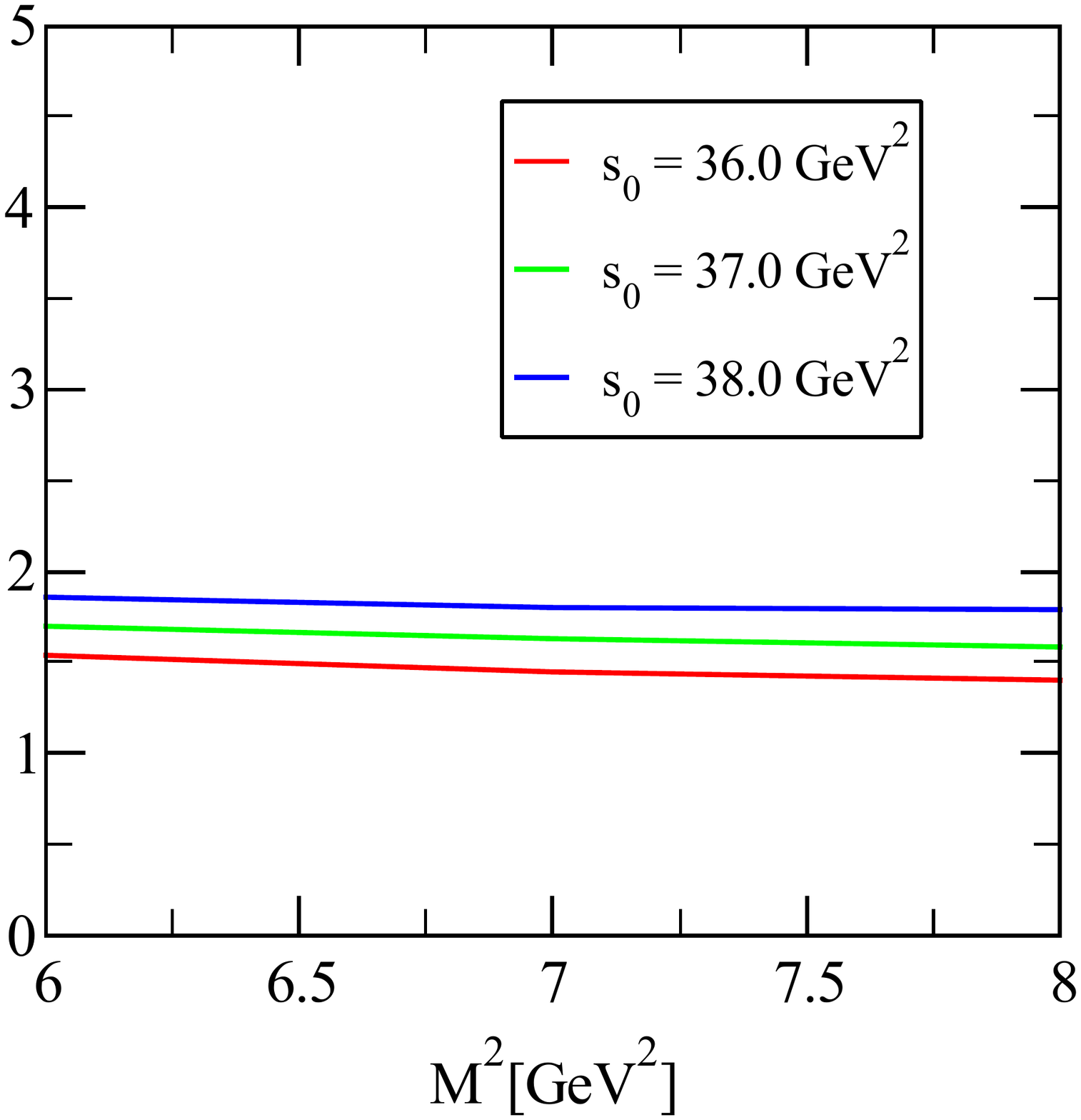}}
 \caption{The dependencies of magnetic moments of  $T_{cc}^{++}$ states on $M^2$ at three different values of $s_0$; (a), (b), (c), (d), (e) and  (f) are represent $J^1_{\mu}$, $J^2_{\mu}$, $J^3_{\mu}$, $J^4_{\mu}$, $J^5_{\mu}$ and $J^6_{\mu}$ states, respectively. }
 \label{Msqfig}
  \end{figure}
  
  \end{widetext}

The physical quantity desired to be calculated in the light-cone sum rules, in our case magnetic moments, includes two extra parameters besides the input parameters given above.
These are the Borel mass parameter $M^2$ and the continuum threshold $s_0$. According to the criteria of the method we used, it is expected that the physical quantities should not change according to these parameters. However, in practice, it is necessary to find the working region where the variation of the calculated physical quantities according to these parameters is minimum. Physical constraints such as pole dominance and convergence of OPE should be applied in order to determine this working region.
Consequently of these restrictions mentioned above, the subsequent working intervals for these extra parameters are given in Table \ref{parameter}. In Fig. \ref{Msqfig}, we depict the dependencies of the magnetic moments versus $M^2$ at fixed values of $s_0$. As one can seen from this figure, the variation of magnetic moments with respect to $M^2$ is quite stable. Though the variation is high compared to $s_0$, this variation remains within the errors of the method used.

\begin{table}[htp]
	\addtolength{\tabcolsep}{10pt}
	\caption{Working regions of the Borel mass parameters and continuum threshold for magnetic moments.}
	\label{parameter}
		\begin{center}
\begin{tabular}{l|c|c|ccc}
	   \hline\hline
   $T_{cc}^{++}$ State& I&  $s_0$~\mbox{[GeV$^2$]}&$M^2$~\mbox{[GeV$^2$]}\\
\hline\hline
$J^1_\mu$&1     &$ 19.5-21.5$        &  $ 4.5-6.5 $\\
$J^2_\mu$&1/2   &$ 20.0-22.0$        &  $ 4.5-6.5 $\\
$J^3_\mu$&0     &$ 21.0-23.0$        &  $ 4.5-6.5 $\\
$J^4_\mu$&1     &$ 33.0-35.0$        &  $ 6.0-8.0 $\\
$J^5_\mu$&1/2   &$ 34.0-36.0$        &  $ 6.0-8.0 $\\
$J^6_\mu$&0     &$ 36.0-38.0$        &  $ 6.0-8.0 $\\
	   \hline\hline
\end{tabular}
\end{center}
\end{table}

After determining all the input parameters required to perform the numerical analysis, the magnetic moment results obtained for the $T_{cc}^{++}$ states using these input parameters are given as follows

\begin{align}\label{MMres}
 \mu_{J^1_\mu} &= 1.21^{+0.54}_{-0.43} \, \mu_N,~~~~~~
 \mu_{J^4_\mu} = 1.76^{+0.25}_{-0.25} \, \mu_N,
 \nonumber\\
 \mu_{J^2_\mu} &= 1.03^{+0.42}_{-0.37} \, \mu_N,~~~~~~
 \mu_{J^5_\mu} = 1.68^{+0.25}_{-0.24} \, \mu_N, 
 \nonumber\\
 \mu_{J^3_\mu} &= 0.97^{+0.37}_{-0.32} \, \mu_N,~~~~~~
 \mu_{J^6_\mu} = 1.63^{+0.23}_{-0.22} \, \mu_N.
\end{align}
Errors in the results in Eq. (\ref{MMres}) are due to uncertainties in all input parameters, numerical values of the parameters in photon distribution amplitudes, as well as extra parameters such as $M^2$ and $s_0$.
In Ref. \cite{Azizi:2021aib}, the magnetic moment of $T^+_{cc}$ state have been acquired within light-cone sum rules by assuming its as diquark-diquark-antiquark and molecular configurations. In the molecular configuration, the result was obtained as $\mu_{T^+_{cc}} = 0.43^{+0.23}_{-0.22} \, \mu_N$. 
Looking at the analysis of the results, it is seen that the obtained values for $T_{cc}^{++}$ states are large. The main reason for this is that the studied tetraquark states have double  electric charge. 
It is to be expected that there will be a two- or three-times difference between single and double charged states, depending on their quark contents, complicated interactions between quarks and gluons etc.. 
The magnitude of the magnetic moment shows its measurability in experiment. From this point of view, it can be said that it is possible to measure the magnetic moments obtained experimentally. Hopefully, our study may attract the lattice QCD and experiment plans in future. 
It is also worth noting that the states have $I=0$ and $I =1$ studied here are predicted to be unable to form bound states when different models are used~( see e.g. Ref.~\cite{Deng:2021gnb,Chen:2021spf,Dai:2021vgf,Karliner:2021wju,Junnarkar:2018twb,Wu:2021rrc,Li:2012ss,Yang:2020fou} and references therein). 
However, we should also state that in some studies in the literature, the spectroscopic parameters of these states were obtained~\cite{Ren:2021dsi,Xin:2021wcr,Albaladejo:2021vln,Braaten:2020nwp,Cheng:2020wxa, Agaev:2018vag,Deng:2018kly,Guo:2021yws}.
In some cases, theoretical explanations of structures for experimentally reported multi-quark states and the availability of theoretical predictions of possible multi-quark states remain problematic. Therefore, more precise experimental investigations of multi-quark states will be required to check the results of theoretical investigations and to develop theoretical models. Doubly-charmed tetraquark molecular states can search without strange, with strange and with double-strange in the $DD\pi $, $DD\gamma $, $DD\pi \pi$,  $DD_s\gamma$, $DD_s\pi\pi$,  $DD_s\pi\pi\pi$, $D_sD_s\gamma$,  $D_sD_s\pi\pi\pi$ invariant mass distributions  at the experimental facilities in future.

In summary, in this work, we estimate magnetic moments of possible doubly-charged  $T_{cc}^{++}$ states within the light-cone sum rules.
 While calculating the magnetic moments of these states, these particles are considered in the molecular picture and they have $J^P = 1^+$ quantum numbers. The magnitude of the magnetic moment shows its measurability in experiment. The magnetic moment results obtained for the $T_{cc}^{++}$ states are large due to the double electric charge. The results obtained in this study can be checked using other theoretical models.
The magnetic moments of the doubly-charmed tetraquark states reveal valuable knowledge about the size and the shape of the hadrons. Determining these parameters is an important step in our interpretation of the hadron properties with respect to quark-gluon degrees of freedom.
It would be exciting to anticipate future experimental efforts that will look for possible $ T_{cc}^{++}$ states and test the findings from the present study.

  \begin{widetext}
  \appendix
  \subsection*{Appendix: Explicit expression for \texorpdfstring{$\Pi^{QCD}$}{}}
  In this appendix, the explicit forms of the analytical expression obtained for $\Pi^{QCD}$ are given as follows
  \begin{align}
   \Pi^{QCD} &= \frac {e_s} {283115520  \, \pi^5} \Bigg[-20 P_ 1 \Big (10 I[0, 4, 2, 
         0] - 30 I[0, 4, 2, 1] + 30 I[0, 4, 2, 2] - 
       10 I[0, 4, 2, 3] - 29 I[0, 4, 3, 0] \nonumber\\
       &+ 58 I[0, 4, 3, 1] - 
       29 I[0, 4, 3, 2] + 28 I[0, 4, 4, 0] - 28 I[0, 4, 4, 1] - 
       9 I[0, 4, 5, 0] + 
       4 \big (28 I[1, 3, 2, 1] - 29 I[1, 3, 2, 2] \nonumber\\
       &+ 
           10 I[1, 3, 2, 3] - 56 I[1, 3, 3, 1] + 29 I[1, 3, 3, 2] + 
           28 I[1, 3, 4, 1]\big)\Big) - 
    4 m_c^2 \Bigg (80 m_s P_ 3 \pi^2 \Big (8 P_ 1 (I[0, 1, 1, 0] \nonumber\\
    &- 
             I[0, 1, 1, 1] - I[0, 1, 2, 0]) - 
          9 \big (I[0, 3, 2, 0] - 2 I[0, 3, 2, 1] + I[0, 3, 2, 2] - 
              2 I[0, 3, 3, 0] + 2 I[0, 3, 3, 1] + I[0, 3, 4, 0] \nonumber\\
              &+ 
              6 I[1, 2, 2, 1] - 3 I[1, 2, 2, 2] - 
              6 I[1, 2, 3, 1]\big)\Big) + 
       40 P_ 1 \Big (-3 I[0, 3, 1, 0] + 5 I[0, 3, 1, 1] - 
          3 I[0, 3, 1, 2] + I[0, 3, 1, 3]\nonumber\\
          &+ 6 I[0, 3, 2, 0] - 
          8 I[0, 3, 2, 1] + 4 I[0, 3, 2, 2] - 
          3 \big (I[0, 3, 3, 0] - I[0, 3, 3, 1] + 3 I[1, 2, 1, 1] - 
              4 I[1, 2, 1, 2] \nonumber\\
              &+ I[1, 2, 1, 3] - 6 I[1, 2, 2, 1] + 
              4 I[1, 2, 2, 2] + 3 I[1, 2, 3, 1]\big)\Big) + 
       27 \Big (2 I[0, 5, 2, 1] - 5 I[0, 5, 2, 2] + 4 I[0, 5, 2, 3] \nonumber\\
       &- 
           I[0, 5, 2, 4] - 4 I[0, 5, 3, 1] + 6 I[0, 5, 3, 2] - 
           2 I[0, 5, 3, 3] + 2 I[0, 5, 4, 1] - I[0, 5, 4, 2] + 
           5 \Big (I[1, 4, 2, 2] \nonumber\\
           &- 2 I[1, 4, 2, 3] + I[1, 4, 2, 4] - 
               2 I[1, 4, 3, 2] + 2 I[1, 4, 3, 3] + 
               I[1, 4, 4, 2]\Big)\Big)\Bigg) + 
    16 m_c \Bigg (20 P_ 1 \Big (4 P_ 3 \pi^2 \big (I[0, 2, 1, 0] \nonumber\\
    &- 
             2 I[0, 2, 1, 1] + I[0, 2, 1, 2] - 2 I[0, 2, 2, 0] + 
             2 I[0, 2, 2, 1] + I[0, 2, 3, 0] - 
             2 (I[1, 1, 1, 0] - 2 I[1, 1, 1, 1] + I[1, 1, 1, 2]\nonumber\\
             &- 
                 2 I[1, 1, 2, 0] + 2 I[1, 1, 2, 1] + 
                 I[1, 1, 3, 0])\big) + 
          m_s \big (I[0, 3, 1, 0] - 3 I[0, 3, 1, 1] + 
              3 I[0, 3, 1, 2] - I[0, 3, 1, 3] \nonumber\\
              &- 2 I[0, 3, 2, 0] + 
              4 I[0, 3, 2, 1] - 2 I[0, 3, 2, 2] + I[0, 3, 3, 0] - 
              I[0, 3, 3, 1] + 
              3 (I[1, 2, 1, 1] - 2 I[1, 2, 1, 2] + I[1, 2, 1, 3] \nonumber\\
              &- 
                  2 I[1, 2, 2, 1] + 2 I[1, 2, 2, 2] + 
                  I[1, 2, 3, 1])\big)\Big) + 
       540 P_ 3 \pi^2 (I[0, 4, 2, 0] - 3 I[0, 4, 2, 1] + 
          3 I[0, 4, 2, 2] - I[0, 4, 2, 3] \nonumber\\
          &- 3 I[0, 4, 3, 0] + 
          6 I[0, 4, 3, 1] - 3 I[0, 4, 3, 2] + 3 I[0, 4, 4, 0] - 
          3 I[0, 4, 4, 1] - I[0, 4, 5, 0] + 
          4 (3 I[1, 3, 2, 1] \nonumber\\
          &- 3 I[1, 3, 2, 2] + I[1, 3, 2, 3] + 
             3 (-2 I[1, 3, 3, 1] + I[1, 3, 3, 2] + I[1, 3, 4, 1]))) + 
       81 m_s \Big (I[0, 5, 2, 0] - 4 I[0, 5, 2, 1] \nonumber\\
       &+ 
           6 I[0, 5, 2, 2] - 4 I[0, 5, 2, 3] + I[0, 5, 2, 4] - 
           3 I[0, 5, 3, 0] + 9 I[0, 5, 3, 1] - 9 I[0, 5, 3, 2] + 
           3 I[0, 5, 3, 3] \nonumber\\
           &+ 3 I[0, 5, 4, 0] - 6 I[0, 5, 4, 1] + 
           3 I[0, 5, 4, 2] - I[0, 5, 5, 0] + I[0, 5, 5, 1] + 
           5 \big (I[1, 4, 2, 1] - 3 I[1, 4, 2, 2] \nonumber\\
           &+ 3 I[1, 4, 2, 3] -
                I[1, 4, 2, 4] - 
               3 (I[1, 4, 3, 1] - 2 I[1, 4, 3, 2] + I[1, 4, 3, 3] - 
                  I[1, 4, 4, 1] + I[1, 4, 4, 2]) \nonumber\\
                  &- 
               I[1, 4, 5, 1]\big)\Big)\Bigg) + 
    405 \big (I[0, 6, 3, 0] - 4 I[0, 6, 3, 1] + 6 I[0, 6, 3, 2] - 
        4 I[0, 6, 3, 3] + I[0, 6, 3, 4] - 3 I[0, 6, 4, 0] \nonumber\\
        &+ 
        9 I[0, 6, 4, 1] - 9 I[0, 6, 4, 2] + 3 I[0, 6, 4, 3] + 
        3 I[0, 6, 5, 0] - 6 I[0, 6, 5, 1] + 3 I[0, 6, 5, 2] - 
        I[0, 6, 6, 0] + I[0, 6, 6, 1] \nonumber\\
        &+ 
        6 (I[1, 5, 3, 1] - 3 I[1, 5, 3, 2] + 3 I[1, 5, 3, 3] - 
            I[1, 5, 3, 4] - 
            3 (I[1, 5, 4, 1] - 2 I[1, 5, 4, 2] + I[1, 5, 4, 3]  \nonumber\\
               &- 
               I[1, 5, 5, 1]+ I[1, 5, 5, 2]) - 
            I[1, 5, 6, 1])\big)\Bigg]\nonumber\\
    &
    +
\frac {e_c} {566231040 \, \pi^5}\Bigg[-16 m_c^2 \Bigg (160 m_s P_ 3 \
\pi^2 \Big (2 P_ 1 \big (I[0, 1, 1, 0] - I[0, 1, 1, 1] - 
             I[0, 1, 2, 0]\big) + 
          99 \big (I[0, 3, 1, 0] \nonumber\\
          &- 2 I[0, 3, 1, 1] + I[0, 3, 1, 2] - 
              2 I[0, 3, 2, 0] + 2 I[0, 3, 2, 1] + 
              I[0, 3, 3, 0]\big)\Big) + 
       27 \big (I[0, 5, 1, 2] - 2 I[0, 5, 1, 3] \nonumber\\
       &+ I[0, 5, 1, 4] - 
          2 I[0, 5, 2, 2] + 2 I[0, 5, 2, 3] + I[0, 5, 3, 2]\big) + 
       5 P1\Big (I[0, 3, 1, 0] - 28 I[0, 3, 1, 1] + 
           37 I[0, 3, 1, 2] \nonumber\\
           &- 10 I[0, 3, 1, 3] - 2 I[0, 3, 2, 0] + 
           30 I[0, 3, 2, 1] - 12 I[0, 3, 2, 2] + I[0, 3, 3, 0] - 
           2 I[0, 3, 3, 1] + 
           6 \big (I[1, 2, 1, 1] \nonumber\\
           &- 6 I[1, 2, 1, 2] + 
               5 I[1, 2, 1, 3] - 2 I[1, 2, 2, 1] + 6 I[1, 2, 2, 2] + 
               I[1, 2, 3, 1]\big)\Big)\Bigg) + 
    96 m_c \Bigg (-40 P_ 1 P_ 3 \pi^2 \Big (I[0, 2, 1, 0] \nonumber
    \end{align}
    \begin{align}
    &- 
          2 I[0, 2, 1, 1] + I[0, 2, 1, 2] - 2 I[0, 2, 2, 0] + 
          2 I[0, 2, 2, 1] + I[0, 2, 3, 0] - 
          2 (I[1, 1, 1, 0] - 2 I[1, 1, 1, 1] + I[1, 1, 1, 2]\nonumber\\
          &- 
              2 I[1, 1, 2, 0] + 2 I[1, 1, 2, 1] + 
              I[1, 1, 3, 0])\Big) + 
       120 m_ 0^2 P_ 3 \pi^2 \Big (I[0, 3, 2, 0] - 2 I[0, 3, 2, 1] + 
          I[0, 3, 2, 2] - 2 I[0, 3, 3, 0] \nonumber\\
          &+ 2 I[0, 3, 3, 1] + 
          I[0, 3, 4, 0] + 6 I[1, 2, 2, 1] - 3 I[1, 2, 2, 2] - 
          6 I[1, 2, 3, 1]\Big) + 
       10 m_s P_ 1 \Big (I[0, 3, 1, 0] - I[0, 3, 1, 1]\nonumber\\
       &- 
          I[0, 3, 1, 2] + I[0, 3, 1, 3] - 2 I[0, 3, 2, 0] + 
          2 I[0, 3, 2, 2] + I[0, 3, 3, 0] + I[0, 3, 3, 1] - 
          3 (I[1, 2, 1, 1] - 2 I[1, 2, 1, 2] \nonumber\\
          &+ I[1, 2, 1, 3] - 
              2 I[1, 2, 2, 1] + 2 I[1, 2, 2, 2] + 
              I[1, 2, 3, 1])\Big) - 
       120 P_ 3 \pi^2 \Big (3 I[0, 4, 1, 0] - 9 I[0, 4, 1, 1] + 
          9 I[0, 4, 1, 2] \nonumber\\
          &- 3 I[0, 4, 1, 3] - 10 I[0, 4, 2, 0] + 
          21 I[0, 4, 2, 1] - 12 I[0, 4, 2, 2] + I[0, 4, 2, 3] + 
          11 I[0, 4, 3, 0] - 13 I[0, 4, 3, 1] \nonumber\\
          &+ 2 I[0, 4, 3, 2] - 
          4 I[0, 4, 4, 0] + I[0, 4, 4, 1] - 
          4 (I[1, 3, 2, 1] - 2 I[1, 3, 2, 2] + I[1, 3, 2, 3] - 
              2 I[1, 3, 3, 1] + 2 I[1, 3, 3, 2]\nonumber\\
              &+ 
              I[1, 3, 4, 1])\Big) + 
       9 m_s \Big (12 I[0, 5, 1, 1] - 36 I[0, 5, 1, 2] + 
           36 I[0, 5, 1, 3] - 12 I[0, 5, 1, 4] - 38 I[0, 5, 2, 1] + 
           77 I[0, 5, 2, 2] \nonumber\\
           &- 40 I[0, 5, 2, 3] + I[0, 5, 2, 4] + 
           40 I[0, 5, 3, 1] - 42 I[0, 5, 3, 2] + 2 I[0, 5, 3, 3] - 
           14 I[0, 5, 4, 1] + I[0, 5, 4, 2] \nonumber\\
           &- 
           5 (I[1, 4, 2, 2] - 2 I[1, 4, 2, 3] + I[1, 4, 2, 4] - 
               2 I[1, 4, 3, 2] + 2 I[1, 4, 3, 3] + 
               I[1, 4, 4, 2])\Big)\Bigg) + 
    15 \Bigg (P_ 1 \Big (-17 I[0, 4, 2, 0] \nonumber\\
    &+ 51 I[0, 4, 2, 1] - 
           51 I[0, 4, 2, 2] + 17 I[0, 4, 2, 3]+ 35 I[0, 4, 3, 0] - 
           70 I[0, 4, 3, 1] + 35 I[0, 4, 3, 2] - 19 I[0, 4, 4, 0] 
           \nonumber\\
           &+ 
           19 I[0, 4, 4, 1] + I[0, 4, 5, 0] - 
           4 (19 I[1, 3, 2, 1] - 35 I[1, 3, 2, 2] + 
               17 I[1, 3, 2, 3] - 38 I[1, 3, 3, 1] + 
               35 I[1, 3, 3, 2]
               \nonumber\\
           &+ 19 I[1, 3, 4, 1])\Big) + 
        2016 m_s P_ 3 \pi^2 \Big (I[0, 4, 2, 0] - 3 I[0, 4, 2, 1] + 
           3 I[0, 4, 2, 2] - I[0, 4, 2, 3] - 3 I[0, 4, 3, 0] + 
           6 I[0, 4, 3, 1] 
           \nonumber\\
           &- 3 I[0, 4, 3, 2] + 3 I[0, 4, 4, 0] - 
           3 I[0, 4, 4, 1] - I[0, 4, 5, 0] + 
           4 \big (3 I[1, 3, 2, 1] - 3 I[1, 3, 2, 2] + 
               I[1, 3, 2, 3] + 
               3 (-2 I[1, 3, 3, 1] 
               \nonumber\\
           &+ I[1, 3, 3, 2] + 
                   I[1, 3, 4, 1])\big)\Big) - 
        18 \Big (2 I[0, 6, 2, 1] - 7 I[0, 6, 2, 2] + 
            9 I[0, 6, 2, 3] - 5 I[0, 6, 2, 4] + I[0, 6, 2, 5] - 
            6 I[0, 6, 3, 1] 
            \nonumber\\
           &+ 15 I[0, 6, 3, 2] - 12 I[0, 6, 3, 3] + 
            3 I[0, 6, 3, 4] + 6 I[0, 6, 4, 1] - 9 I[0, 6, 4, 2] + 
            3 I[0, 6, 4, 3] - 2 I[0, 6, 5, 1] + I[0, 6, 5, 2] 
            \nonumber\\
           &+ 
            6 \big (I[1, 5, 2, 2] - 3 I[1, 5, 2, 3] + 
                3 I[1, 5, 2, 4] - I[1, 5, 2, 5] - 
                3 (I[1, 5, 3, 2] - 2 I[1, 5, 3, 3] + I[1, 5, 3, 4] - 
                   I[1, 5, 4, 2]
                   \nonumber\\
           &+ I[1, 5, 4, 3]) - 
                I[1, 5, 5, 2]\big)\Big)\Bigg)\Bigg]\nonumber\\
                &
                +m_0^2 e_s \Bigg( 25  I_4[\mathcal A]\, I[0, 5, 3, 0]+ 2 \, I_3[\mathcal T_1] I[0, 5, 4, 0]+578\, f_{3\gamma}\, \pi^2\, I_2[\mathcal V]\, I[0, 4, 4, 0] + \chi \Big(476\, I_1[\mathcal V]\, I[0, 6, 4, 0] 
 \nonumber\\
&
+ 39  I_3[\mathcal S]\, I[0, 6, 3, 0]+ 8\, I_4[\mathcal T_4] I[0, 6, 2, 0]\Big)\Bigg),
\label{app}
  \end{align}
where $P_1 =\langle g_s^2 G^2\rangle$, $P_3 =\langle \bar s s \rangle$ are gluon and  s-quark condensates, respectively. We should also mention that, in the Eq. (\ref{app}), for simplicity we have only presented the terms that give substantial contributions to the numerical values of the observables under investigation and ignored to present many higher dimensional operators though they have been considered in the numerical calculations.

The functions~$I[n,m,l,k]$, $I_1[\mathcal{F}]$,~$I_2[\mathcal{F}]$,~$I_3[\mathcal{F}]$ and~$I_4[\mathcal{F}]$ are
defined as:
\begin{align}
 I[n,m,l,k]&= \int_{4m_c^2}^{s_0} ds \int_{0}^1 dt \int_{0}^1 dw~ e^{-s/M^2}~
 s^n\,(s-4m_c^2)^m\,t^l\,w^k,\nonumber\\
 I_1[\mathcal{F}]&=\int D_{\alpha_i} \int_0^1 dv~ \mathcal{F}(\alpha_{\bar q},\alpha_q,\alpha_g)
 \delta'(\alpha_ q +\bar v \alpha_g-u_0),\nonumber\\
  I_2[\mathcal{F}]&=\int D_{\alpha_i} \int_0^1 dv~ \mathcal{F}(\alpha_{\bar q},\alpha_q,\alpha_g)
 \delta'(\alpha_{\bar q}+ v \alpha_g-u_0),\nonumber\\
   I_3[\mathcal{F}]&=\int D_{\alpha_i} \int_0^1 dv~ \mathcal{F}(\alpha_{\bar q},\alpha_q,\alpha_g)
 \delta(\alpha_ q +\bar v \alpha_g-u_0),\nonumber\\
   I_4[\mathcal{F}]&=\int D_{\alpha_i} \int_0^1 dv~ \mathcal{F}(\alpha_{\bar q},\alpha_q,\alpha_g)
 \delta(\alpha_{\bar q}+ v \alpha_g-u_0),\nonumber
 \end{align}
 where $\mathcal{F}$ stands for the corresponding photon DAs.

   \end{widetext}
 \bibliography{Tcc-Molecule}

\end{document}